\documentclass[twocolumn,superscriptaddress,nofootinbib,preprintnumbers,showpacs,tightenlines,notitlepage]{revtex4}
\usepackage[latin1]{inputenc}
\usepackage{graphicx}
\usepackage{amssymb}
\usepackage{color}
\usepackage{float}
\usepackage{amsmath}
\usepackage{amsfonts}
\usepackage{dcolumn}
\usepackage{hyperref}

\begin{document}

\title{DGP black holes on the brane}

\author{Radouane Gannouji}
\affiliation{Department of Physics, Faculty of Science, Tokyo
University of Science, 1-3, Kagurazaka, Shinjuku-ku, Tokyo 162-8601,
Japan}

\begin{abstract}
We find an exact solution on the brane for static black hole in the DGP model. In the appropriate limit we recover the 2 known solutions, the Schwarzschild and the Reissner-Nordstr\"om solutions with tidal charge. The solution has 2 branches, which correspond asymptotically to de-Sitter or flat Universe. Also we study the linear stability of the solutions. We find that the Regge-Wheeler and the Zerilli potential are positive. Finally the dipole perturbation is derived which corresponds to a linearization of a rotating black hole solution on the brane.
\end{abstract}

\pacs{04.50.Kd,04.50.Gh,04.20.Jb}

\maketitle


Extra dimensions provide an approach to modify gravity without abandoning the form of the action proposed in the Einstein's general relativity. From a phenomenological point of view we can avoid constraints coming from standard model observations, by considering a brane world scenario, that is, we are living in a hypersurface (3-D) in a higher dimensional spacetime. From the theoretical point of view, string theory predicts a boundary layer, a brane, on which edges of open strings stand \cite{Polchinski:1995mt}. The possibility that we may be living in a brane generates many questions as how gravity looks like. Also in an attempt to solve the much debated hierarchy problem, various problems are studied, but also in order to understand the cosmology, such as inflation and Dark Energy. 
In this contribution to consequences of the brane-world in 4-D, we study one of the most famous model, DGP \cite{Dvali:2000hr}.

We know that the physics of black hole and gravitational collapse is complicated, especially because of the matter localized to the brane, while the gravitational field can access the extra dimension, and also because of the nonlocal effects of the bulk into the surface. Part of the problem would be to find a solution on the brane. We can not necessarily embed this solution into a bulk but some information of the global solution can be understood, some intuition can be developed. The solution can be smoothly continued into the bulk via the ADM formalism, where the solution on the brane can be considered as an initial data. At least a local solution of the bulk exist, even if the global solution is not guaranteed.

The model is defined as an empty five-dimensional space (not necessarily Minkowsky) and all the energy-momentum is localised on the four-dimensional brane. The theory is described by the following action in the vacuum
\begin{align}
\label{eq:action}
\mathcal{S}=M_{(5)}^3\int {\rm d^5x} \sqrt{-g}R
+\frac{M_{(4)}^2}{2}\int {\rm d^4x}\sqrt{-h} R,
\end{align}
where $(g,h)$ are respectively the metric of the bulk and the brane
and $R$ the intrinsic curvature in 5-D and 4-D respectively. The Gibbons-Hawking term is implicit in (\ref{eq:action}).

Variation of the action gives the following equations of motion

\begin{align}
\text{Bulk equation:}\quad  R_{\mu\nu}&=0,\\
\text{Brane equation:}\quad G_{\mu\nu}&=\frac{1}{r_c}\Bigl(K_{\mu\nu}-Kh_{\mu\nu}\Bigr).
\end{align}

where $r_c=M_{(4)}^2/2M_{(5)}^3$ is the crossover scale that governs the transition between four-dimensional behaviour and five-dimensional behaviour.

Following \cite{Shiromizu:1999wj} and \cite{Kofinas:2002gq} we can rewrite an equation on the brane with the metric $h$ only. For this we define the tensor \cite{Kofinas:2002gq}

\begin{align}
L_{\mu\nu}=K_{\mu\nu}-\frac{K}{2}h_{\mu\nu}+\frac{1}{2r_c}h_{\mu\nu},
\end{align}

which gives on the brane 

\begin{align}
\label{Eq:brane}
G_{\mu\nu}+\frac{3}{2r_c^2}h_{\mu\nu}=\frac{1}{r_c}\Bigl(L_{\mu\nu}+\frac{L}{2}h_{\mu\nu}\Bigr),
\end{align}

and $L$ is solution of the following algebraic equation (Gauss equation)

\begin{align}
\label{Eq:L}
L_\mu^{~\alpha}L_{\alpha\nu}-\frac{L^2}{4}h_{\mu\nu}+\frac{3}{4r_c^2}h_{\mu\nu}=-E_{\mu\nu},
\end{align}

where $L$ is the trace of the tensor and $E$ is the electric part of the Weyl tensor.


In the following, we will focus on static spherically solutions in the vacuum of the form

\begin{align}
\label{eq:metric}
{\rm d}s^2=-A(r){\rm d}t^2+\frac{{\rm d}r^2}{B(r)}+r^2{\rm d}\Omega^2.
\end{align}

Therefore we can decompose irreducibly the tensor $E$ with respect to a $4$-velocity field $u^\mu$ \cite{Maartens:2000fg,Dadhich:2000am}

\begin{align}
\label{Eq:E}
E_{\mu\nu}=\rho(r)\Bigl(u_{\mu}u_{\nu}+\frac{1}{3}q_{\mu\nu}\Bigr)+P(r)\Bigl(r_{\mu}r_{\nu}-\frac{1}{3}q_{\mu\nu}\Bigr),
\end{align}

where $q_{\mu\nu}=h_{\mu\nu}+u_{\mu}u_{\nu}$ projects orthogonal to the timelike vector. Here $(\rho,P)$ are respectively an effective energy density and anisotropic stress on the brane arising from the $5$-D gravitational field.

It is easy to see from (\ref{Eq:brane},\ref{eq:metric}) that $L$ is diagonal, hence we write $L^\mu_{~\nu}=\text{diag} \Bigl(L_0,L_1,L_2,L_3\Bigr)$.

From the equations (\ref{Eq:L},\ref{Eq:E}) we have the following algebraic equations

\begin{align}
\label{Eq:L1}
L_1&=\pm\sqrt{L_0^2-\frac{2}{3}(2\rho+P)},\\
\label{Eq:L2}
L_2&=L_3=\pm\sqrt{L_0^2-\frac{1}{3}(4\rho-P)},
\end{align}

where the equation $L_2=L_3$ comes from the Eq.(\ref{Eq:brane}).

Also from the brane equation (\ref{Eq:brane}), we have

\begin{align}
-A\frac{\rm d}{\rm d r}\Bigl(\frac{B}{A}\Bigr)=\frac{r}{r_c}\Bigl(L_1-L_0\Bigr).
\end{align}

Hereafter, we will assume that radial photons should experience no acceleration, the velocity of light in the radial direction should remain constant. Therefore we have \cite{Dadhich:2012pd} $A=B$ which implies $L_1=L_0$, hence from (\ref{Eq:L1}) we have $2\rho+P=0$. This constraint between the density and the pressure is the same than in the absence of induced curvature term \cite{Dadhich:2000am}.
 
Finally, we have from (\ref{Eq:L}) 

\begin{align}
4L_0^2+2P\pm 8L_0\sqrt{L_0^2+P}=\frac{3}{r_c^2}.
\end{align}

Following \cite{Kofinas:2002gq}, we define $v=2\pm\sqrt{L_0^2+P}/L_0$. Then, it is straightforward to check that

\begin{align}
\label{Eq:L02}
L_0^2&=\frac{3}{2r_c^2(v^2-3)},\\
\label{Eq:P}
P&=\frac{3}{2r_c^2}\frac{(v-1)(v-3)}{v^2-3}.
\end{align}

We see that we need $v^2>3$. 

The only undetermined function is $v$, all the other quantities as $(\rho,P,K_{\mu\nu})$ are related to $v$.

Considering now the Bianchi identity

\begin{align}
\nabla_\mu L^\mu_{~\nu}+\frac{1}{2}\nabla_\nu L=0,
\end{align}

we can close the system of equations and get an equation for $v$

\begin{align}
\frac{{\rm d}v}{{\rm d} r}+\frac{2}{3r}(v-3)(v^2-3)=0,
\end{align}

which gives

\begin{align}
\label{Eq:v}
\frac{r^4}{r_c^2}=Q^2 \frac{|v-\sqrt{3}|^{(\sqrt{3}+1)/2}}{|v+\sqrt{3}|^{(\sqrt{3}-1)/2}|v-3|},
\end{align}

where $Q$ is an integration constant.

The coefficient $r_c^2$ is fixed in order to recover the result $P\simeq 1/r^4$ \cite{Dadhich:2000am} in the limit $r_c\rightarrow 0$, which would corresponds to the Reissner-Nordstr\"om solution on the brane with tidal charge.

The eq.(\ref{Eq:v}) gives $v(r)$ which from (\ref{Eq:L02}) gives $L_0(r)$, therefore we can solve the final equation (\ref{Eq:brane})

\begin{align}
\frac{rB'(r)+B(r)-1}{r^2}=-\frac{3}{2r_c^2}\pm \frac{\sqrt{3/2}}{r_c^2}\frac{v}{\sqrt{v^2-3}},
\end{align}

where the sign $\pm$ is because of the sign of $L_0$. 

We have found 3 different solutions depending on the range of $v$. In the first solution we have $v<-\sqrt{3}$, the second solution corresponds to $\sqrt{3}<v<3$ and the last to $v>3$. Accordingly the range for $r$ will be respectively $r>\sqrt{Qr_c}$, $r>0$ and $r>\sqrt{Qr_c}$. The second and the third solutions are identical except the range for $r$, hence we keep only the second solution which covers the full spacetime (brane). The first solution do not cover the full spacetime $r>\sqrt{Qr_c}$ and therefore can't describe a black hole. This solution will not be studied in this paper. Hence we have 2 branches of the solution

\begin{align}
\label{metric}
A\equiv B=1-\frac{2m}{r}-\frac{r^2}{2r_c^2}\pm \frac{r^2}{2r_c^2}f(v),
\end{align}

where $m$ is an integration constant, $v$ is solution of the algebraic equation (\ref{Eq:v}) and $f$ can be written in terms of Gauss's hypergeometric function,

\begin{widetext}
\begin{align}
f(v)=\sqrt{6}\frac{v-2}{\sqrt{v^2-3}}-\frac{4\sqrt{2}(\sqrt{3}-1)(v-3)^2}{5(v-\sqrt{3})^{3/2}(v+\sqrt{3})^{1/2}}~_2F_1\Bigl[1,\frac{3(3-\sqrt{3})}{8},\frac{9}{4},(\sqrt{3}-1)\frac{v-3}{v-\sqrt{3}}\Bigr].
\end{align}
\end{widetext}

$f$ is a concave down monotonically increasing function of $r$. It is negative for $r\lesssim 0.78 \sqrt{Qr_c}$ and positive otherwise and $\lim_{r\rightarrow \infty} f=1$. 

The horizon structure of the black hole on the brane depends on the branch considered. The negative branch (negative sign in the equation (\ref{metric})) has a black hole horizon and a cosmological horizon because of the de-Sitter structure at large distances, while the positive branch, which is asymptotically flat, has a single horizon. 


It is interesting to study the asymptotic behaviour of this solution. We have at large distances

\begin{align}
\label{linear}
f(r)= 1-\frac{2q^2r_c^2}{r^4}+\frac{2q^4r_c^4}{5r^8}-\frac{4q^6r_c^6}{9r^{12}}+O(\frac{1}{r^{16}}),
\end{align}

where we have redefined the integration constant $q=\sqrt{3/2}(3-\sqrt{3})^{(\sqrt{3}-1)/4}(3+\sqrt{3})^{-(\sqrt{3}+1)/4} Q$.

We see that only the positive branch has a smooth limit as $r_c\rightarrow 0$ (Randall-Sundrum model limit) and as such we will refer to as the RS branch. In contrast the negative branch is not smooth as $r_c\rightarrow 0$ and represents a distinct new feature of DGP, the DGP branch also known as the self-accelerating branch. The RS branch converges to $A=1-2m/r-q^2/r^2$, it is not a Reissner-Nordstr\"om spacetime. In fact the tidal charge has always the same sign and it is physically more natural for a brane solution \cite{Dadhich:2000am}, the tidal charge strengthens the gravitational field. This is why our solution do not have a Cauchy horizon, even in the limit $r_c\rightarrow 0$. Also as $r_c\rightarrow \infty$, we recover the Schwarzschild solution ($1-2m/r$) for the 2 branches, the solution is the same than in Einstein theory, there is no vDVZ \cite{vanDam:1970vg,Zakharov:1970cc} discontinuity. The continuity of the theory is restored because the nonlinear effects were taken into account while  we would conclude to a discontinuity if we use the linearized solution at large distances (\ref{linear}).



At small distances

\begin{align}
f(r)\simeq  -\frac{\alpha}{W^{3 (1+\sqrt{3})/8}}+\frac{\beta}{\sqrt{W}}+\frac{\gamma}{W^{(3\sqrt{3}-5)/8}},
\end{align}

where $(\alpha,\beta,\gamma)$ are 3 positive constants 

\begin{align}
\alpha &=\frac{3 \left(2 \sqrt{3}-3\right)^{\frac{3}{8} \left(\sqrt{3}-3\right)} \Gamma \left(\frac{5}{4}\right) \Gamma \left(\frac{1}{8} \left(3 \sqrt{3}-1\right)\right)}{\sqrt{45+26 \sqrt{3}} ~\Gamma
   \left(\frac{3}{8} \left(3+\sqrt{3}\right)\right)}\simeq 0.76,\nonumber\\
\beta &=\frac{1}{13} \sqrt{90+\frac{111 \sqrt{3}}{2}}\simeq 1.05,\nonumber\\
\gamma &=\frac{3 \sqrt[4]{3} \left(2 \sqrt{3}-3\right)^{\frac{3 \sqrt{3}}{8}-\frac{1}{8}} \Gamma \left(\frac{9}{4}\right) \Gamma \left(\frac{1}{8} \left(3 \sqrt{3}-1\right)\right)}{5 \sqrt{2} \Gamma
   \left(\frac{3}{8} \left(3+\sqrt{3}\right)\right)}\simeq 0.77,\nonumber
\end{align}

and $W$ is the Lambert function at the point $3^{\sqrt{3}/2}4^{1-\sqrt{3}}(2-\sqrt{3}) \Bigl(\frac{r}{\sqrt{qr_c}}\Bigr)^{4(\sqrt{3}-1)}$.

The dominant contribution is given by the first term

\begin{align}
f(r)\simeq -\delta \frac{(r_c q)^{3/2}}{r^3},
\end{align}

with 

\begin{align}
\delta=2 \sqrt{2}\frac{\Gamma \left(\frac{5}{4}\right) \Gamma \left(\frac{1}{8} \left(3 \sqrt{3}-1\right)\right)}{3^{3/8}\Gamma \left(\frac{3}{8} \left(3+\sqrt{3}\right)\right)}\simeq 3.11\nonumber.
\end{align}

We see that even in the case of a massless black hole, we have a "mass-term" because of the fifth dimension. Hence we recover a standard result; even for a massless black hole, the behaviour of the solution is $1/r$ at small distances and $1/r^2$ at large distances.

\begin{figure}
\includegraphics[scale=0.75]{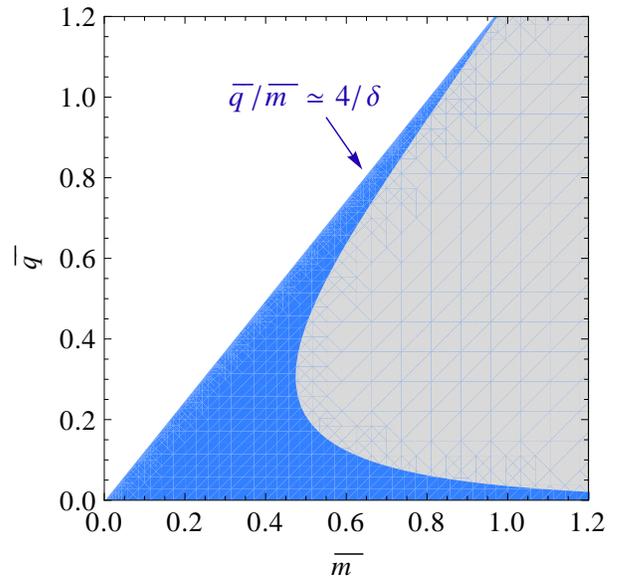}
\caption{Existence of the black hole for the DGP branch: The blue part represents the range of the parameters $(\bar{m},\bar{q})$ for which the black hole has 2 horizons. In the grey region, the metric is always negative and the white region corresponds to a black hole with only a cosmological horizon, because of the negativity of the effective mass, hence inside the cosmological horizon the solution would be a naked singularity.}
\label{fig:f}
\end{figure}

In order to keep the effective mass positive, in the DGP branch, we impose $\bar{q}<\frac{4}{\delta}\bar{m}$, where $(\bar{m}=m/\sqrt{q r_c},\bar{q}=q/r_c)$. The existence of the black hole is constrained in the Fig.(\ref{fig:f}). From which we see that for a fixed parameter $\bar q$, the mass of the black hole has an upper bound but also a lower bound. We would have a naked singularity for lightest black holes, this can be seen as an instability of the branch.

The RS branch is much simpler, we have a black hole for all positive parameters $(\bar{m},\bar{q})$, also these parameters play the same role, they increase the position of the horizon and hence its entropy. At large distances, the Newtonian potential is dominant. Depending on the parameters, the situation can be the same for all distances, the solution will be very close to the Schwarzschild solution, except for large values of $\bar{q}$ where the mass of the black hole will be renormalized at small distances. But in the case of the DGP branch, we do not have the same behaviour at large and small distances, hence we have a new distance scale $r_\star$ dubbed the Vainshtein radius

\begin{align}
r_\star\simeq (mr_c^2)^{1/3}.
\end{align}

As we said previously, the mass of the black hole is bounded from bellow. The existence of the horizon is constrained by $m> \bar q^2 r_c^2/r_\star\simeq \bar q^2 10^{28} M_{\odot}$ if we assume the Vainshtein radius of the order galaxy scale and the crossover scale of the order Hubble scale. A stellar black hole exists if $\bar q<10^{-14}$. Otherwise it will be a naked singularity.

The extrinsic curvature and therefore the curvature constant can be easily derived from (\ref{Eq:L02},\ref{Eq:P}) 

\begin{align}
R=\frac{3}{r_c^2}\Bigl(2\mp\sqrt{6}\frac{v-1}{\sqrt{v^2-3}}\Bigr),
\end{align}
which is singular at $r=0$. Also we can see that $R>0$ for the DGB-branch and converges to $12/r_c^2$, while we have $R<0$ for the RS-branch and it goes to zero at infinity. A non-vanishing curvature outside the source leads usually to screening mechanism and we have shown previously the absence of the vDVZ discontinuity. On the physical stability of the solutions, it is interesting to study the violation of the energy conditions if we consider the tensor $K_{\mu\nu}-Kh_{\mu\nu}$ as a source term, 
and the positivity of the gravitational mass of this spacetime. These particular problems should be addressed separately.\\

In order to study the linear stability of this solution, we follow the Regge-Wheeler formalism \cite{Regge:1957td,Zerilli:1970se} and we decompose the metric perturbations according to their transformation properties under two-dimensional rotations. They are classified depending on the transformation properties under parity, namely odd (axial) and even (polar). Using the Regge-Wheeler, and Zerilli gauge, one obtains two distinct perturbations : odd and even perturbations.

For $\ell > 1$, the equation of perturbations takes the form

\begin{align}
\frac{d^2}{d t^2}\psi-\frac{d^2}{dr^*}\psi+V \psi=S,
\end{align}

where $S$ is the perturbation of the source term, $r^*$ is the tortoise coordinate defined as $dr^{*2}=dr^2/A$, $\psi$ is a function of the metric perturbations and $V$ is the Regge-Wheeler potential or the Zerilli potential in the respective case.

We have for the Regge-Wheeler potential

\begin{align}
V_{RW}=&\frac{A \left(\lambda+2 A-r A' \right)}{r^2}-\frac{12 \rho  r^2 A}{6 r^2+r_c^2 \left[(r^2 A)''-2\right]},\\
=&A\frac{\lambda-1+3A}{r^2}+\frac{3A}{2r_c^2}\Bigl(1\mp \sqrt{\frac{6}{v^2-3}}\Bigr),
\end{align}

where $\lambda=(\ell-1)(\ell+2)$.

The last form is useful in order to see that the potential is always positive for the 2 branches. The first term of the potential is the standard term in 4 dimensional and the last term comes from the 5 dimensional effects. We do not write here the Zerilli potential because it is much more complicated but as in General Relativity, the graph of the Zerilli potential is similar than the Regge-Wheeler potential for the same parameters.

The positivity of the potential indicates the stability of the spacetime under linear perturbations \cite{Vishveshwara:1970cc} for the 2 branches. It is important to notice that we didn't considered the source term in order to study the stability of the theory. In this case, the source term is much more complicated than in General Relativity. In fact it is not localized, the source is function of the electric part of the Weyl tensor which is a non local term.

For the dipole perturbation $\ell=1$, we can write $h_{t\phi}=\beta(r)\sin^2(\theta)$ because the time dependant term can be removed via the gauge freedom. The equation for $\beta$ takes the form
\begin{align}
\label{Hill}
\frac{\beta''}{\beta} =\frac{2}{r^2}\pm \sqrt{\frac{3}{2}}\frac{v-3}{r_c^2 A \sqrt{v^2-3}},
\end{align}
where we have neglected the source term. The solution of this equation represents the spacetime around a slowly rotating black hole on the brane for the DGP model. In the case where $r_c\rightarrow \infty$ which corresponds to the GR limit, we recover the standard result, the Kerr metric at the first order of perturbations.
\begin{align}
h_{t\phi}=-\frac{J}{r}\sin^2(\theta),
\end{align}
where $J$ is identified to the angular momentum and we have gauged away the unphysical term proportional to $r^2$. 

The Eq.(\ref{Hill}) has a Hill type form, hence the evolution will depend on the sign of the RHS of the equation. In order to integrate the system, we assume a Kerr form of the metric at small distances. For the DGP branch, we have an angular speed $\Omega=-\beta(r)/r^2$ which decreases with the radial distance but contrary to General Relativity, $\Omega=0$ at a finite distance, smaller than the cosmological horizon. After this critical point, the frame-dragging is occurring in the opposite direction. In the case of the RS branch, the angular speed depends on the parameters $(\bar m,\bar q)$. For small $\bar q$ we recover Kerr solution. Once we increase this parameter, $\Omega$ goes to a finite value at infinity, and finally when $\bar q>\bar m$ the angular speed increases after decreasing and converge also to a finite value. 

In conclusion, we have derived an exact black hole solution on the brane for the DGP model. This solution recovers the standard results at small and large distances but also covers the intermediate regime which was not known. The 2 branches depend on 3 parameters, the mass of the black hole, the tidal charge and the cross scale parameter. We have shown than if we do not consider the perturbations of the bulk (the source term), the solutions are stable under linear perturbations. 
It would be also interesting to see how the horizon will be affected by the full solution and how the quasinormal modes of the black hole are modified. Finally we have shown from the dipole perturbation, an approximation of a possible rotating black hole on the brane. For the DGP branch, the metric $g_{t\phi}$ decreases to zero in a finite range of the radial distance, while for larger distances, the angular speed is opposite. Whereas $g_{t\phi}$ converges to a finite value at infinity for the RS branch. Contrary to General Relativity, the dragging do not disappears completely at large distances.

\begin{acknowledgments}
It is a pleasure to thank N. Dadhich and M. Sami for useful discussions. I thank also H. Nandan and R. Ul Haq Ansari for their initial collaboration on this work. This research is supported by the Grant-in-Aid for Scientific Research Fund of the JSPS No. 10329.
\end{acknowledgments}

\end{document}